\documentclass[sigconf,screen,nonacm]{acmart}

\usepackage[utf8]{inputenc}
\usepackage{subcaption}
\usepackage{hhline}
\definecolor{codegreen}{rgb}{0,0.6,0}
\definecolor{codegray}{rgb}{0.5,0.5,0.5}
\definecolor{codepurple}{rgb}{0.58,0,0.82}
\definecolor{backcolour}{rgb}{1,0.973,0.906}
\definecolor{mygray}{gray}{0.96}

\usepackage{listings}
\lstdefinestyle{mystyle}{
  backgroundcolor=\color{mygray},
  commentstyle=\color{codegreen},
  keywordstyle=\color{magenta},
  stringstyle=\color{codepurple},
  basicstyle=\footnotesize,
  breakatwhitespace=false,         
  breaklines=true,                 
  captionpos=b,                    
  keepspaces=true,                 
  numbersep=5pt,                  
  showspaces=false,                
  showstringspaces=false,
  showtabs=false,                  
  tabsize=2
}
\usepackage{paralist}
\usepackage{graphicx}
\usepackage{hyperref}
\usepackage{balance}
\usepackage{multirow}
\usepackage[noabbrev,nameinlink]{cleveref}

\setcopyright{acmcopyright}
\copyrightyear{2022}
\acmYear{2022}

\acmConference[WiseML '22]{$15^{th}$ ACM Conference on Security and Privacy in Wireless and Mobile Networks}{May 19, 2022}{San Antonio, Texas, USA}

\title{Can You Hear It? Backdoor Attacks via Ultrasonic Triggers}

\author{Stefanos Koffas}
\affiliation{%
  \institution{Delft University of Technology}
  \country{The Netherlands}
}
\author{Jing Xu}
\affiliation{%
  \institution{Delft University of Technology}
  \country{The Netherlands}
}
\author{Mauro Conti}
\affiliation{%
  \institution{University of Padua}
  \country{Italy}
}
\author{Stjepan Picek}
\affiliation{%
    \institution{Radboud University}
}
\affiliation{%
  \institution{Delft University of Technology}
  \country{The Netherlands}
}


\begin{document}


\begin{abstract}


This work explores backdoor attacks for automatic speech recognition systems where we inject inaudible triggers. By doing so, we make the backdoor attack challenging to detect for legitimate users, and thus, potentially more dangerous. We conduct experiments on two versions of a speech dataset and three neural networks and explore the performance of our attack concerning the duration, position, and type of the trigger. Our results indicate that less than 1\% of poisoned data is sufficient to deploy a backdoor attack and reach a 100\% attack success rate. We observed that short, non-continuous triggers result in highly successful attacks. However, since our trigger is inaudible, it can be as long as possible without raising any suspicions making the attack more effective. Finally, we conducted our attack in actual hardware and saw that an adversary could manipulate inference in an Android application by playing the inaudible trigger over the air.

\end{abstract}

\maketitle

\section{Introduction}
\label{sec:sound-recognition}

Automatic speech recognition (ASR) has gained much attention in recent years as it can be a very efficient form of communication between people and machines. 
Voice assistants like Google Assistant and Amazon Alexa have already shown the accessibility, efficiency, and high recognition accuracy of ASR.
As researchers devote much of their effort to improving the performance of ASR systems, there are also works trying to understand how vulnerable is the ASR system to adversarial attacks like backdoor attacks.

Backdoor attacks represent a serious threat to neural networks. A backdoored model misclassifies the trigger-embedded inputs into an attacker-chosen target label while performing normally on benign inputs. 
A backdoor attack could occur in a voice-controlled device (mobile phone or an IoT device like Alexa) that uses cloud-based inference, a pre-trained model, or transfer learning~\cite{badnets-evaluating-backdoor-attacks-on-dnns}.
To protect systems against such threats, it is important to understand them and investigate in what scenarios they pose the biggest risk. An intuitive answer is in scenarios where the trigger cannot (or is difficult to) be noticed by legitimate users.

Liu et al. implemented a backdoor attack in ASR by injecting background noise to the original audio sample and retraining the model to recognize the stamped audio as a specific word~\cite{trojaning-attack-on-nns}. Zhai et al. designed a backdoor attack against speaker verification methods, which applies a clustering-based attack scheme where poisoned samples from different clusters will contain different triggers~\cite{backdoor-attack-against-speaker-verification}. Xu et al. backdoored the ASR system by generating a random sequence of data points for the trigger~\cite{meta-neural}. All these works inject unintelligible but audible triggers into the training dataset. As a result, humans can understand the differences between the trigger-embedded and the clean inputs. 

In this work, we consider backdoor attacks on ASR systems using an inaudible trigger. It is intuitive that the attacker poses a more severe threat to the system with inaudible triggers that are unnoticeable by humans. To the best of our knowledge, there is no study exploring backdoor attacks with inaudible triggers in speech recognition systems. In this paper, we seek to bridge this gap. We did not exploit existing nonlinearities in the sound processing pipeline~\cite{dolphin-attack} as our goal is to shed light on new risks that can be introduced when a high sampling rate is used. While such sampling rates are not recommended for ASR systems~\cite{google-speech-to-text}, they are still allowed by public speech-to-text APIs~\cite{google-speech-to-text,microsoft-speech-to-text}.
Our main contributions are:
\begin{itemize}
    \item We propose a backdoor attack on a speech recognition system using an ultrasonic pulse as the trigger.
    \item We systematically evaluate the impact of various trigger characteristics on the performance of backdoor attacks on ASR. 
    \item We use two benchmark datasets and three neural networks to explore the performance of backdoor attacks.
    \item We show that our attack is effective with real hardware by attacking an Android application.
\end{itemize}
To foster reproducibility, our code is publicly available~\footnote{\url{https://anonymous.4open.science/r/ultrasonic_backdoor-FBF3}}.

\section{Preliminaries}
\label{sec:background}


\subsection{Automatic Speech Recognition (ASR)}

Computers use automatic speech recognition (ASR) to understand human speech and translate it into text. Modern ASR systems use deep learning for their training. The audio files used for training are 1-dimensional vectors with $N$ elements corresponding to analog audio signals sampled to a specific sampling rate. 

These vectors enclose information like noise and voice color, which is not important for speech classification. Usually, a pre-processing step discards all this information and calculates more appropriate input features like spectrogram images~\cite{trojaning-attack-on-nns}, or the
Mel-Frequency Cepstral Coefficients (MFCCs). We use the MFCCs, which are 1) rather accurate as they emulate the functionality of the human vocal system~\cite{deep-learning-for-nlp-and-speech} and 2) widely used~\cite{meta-neural,adversarial-detection-by-classification}.

\subsection{Backdoor Attacks}

In a backdoor attack, the adversary aims to train a neural network that correctly solves the desired task on expected data but exhibits malicious behavior once presented with a certain trigger~\cite{badnets-evaluating-backdoor-attacks-on-dnns}. 
We use the data poisoning attack, where the adversary adds to the original dataset poisoned examples (examples containing a trigger and labeled as the target) so to force the model trained on this dataset to behave incorrectly~\cite{badnets-evaluating-backdoor-attacks-on-dnns}.

Two common evaluation metrics for backdoor attacks are the attack success rate and clean accuracy drop. The attack success rate is the fraction of successfully triggered backdoors over a set of poisoned inputs. Since both automatic speech recognition and attack success rate have the same abbreviation (ASR), we will use the nonstandard abbreviation ASRT for the attack success rate. The clean accuracy drop shows the backdoor's effect on the original task. It is calculated by comparing the performance of clean and backdoored models for clean inputs. We compute both metrics on the entire test set.


\section{Inaudible Backdoor Attack}
\label{sec:inaudible}

%



\subsection{Trigger}
\label{ssec:trigger}

To the best of our knowledge, we are the first to experiment with a stealthy trigger in the inaudible range (> 20kHz). Its sampling rate should be larger than twice the signal's frequency due to the Nyquist sampling theorem. 
We select the trigger's sampling rate to be 44.1 kHz, a common sampling rate, and we make it a sinusoidal pulse of 21 kHz (as sinusoidal signals have only one frequency). 

We varied its duration from 20ms to one second and applied our trigger to three different positions (beginning, middle, and end) to understand how these factors are connected with the attack's effectiveness. Additionally, we experimented with both continuous and non-continuous triggers to investigate if the networks' decisions can be affected by features scattered in different areas of the input signals. To create a non-continuous trigger of $x$ ms, we distribute in every 1-second audio sample five triggers of $\frac{x}{5}$ms. Our initial exploration gave good results with five triggers, but the best choice will be further investigated in the future.

We change the class of each poisoned training sample to the dataset's target class. We chose different target classes for each dataset version (see ~\Cref{ssec:dataset}) to verify the attack's effectiveness in different settings. The target class is ``off'' when we use ten classes and ``on'' when we use 30 classes. 
We poison the first $N$ samples of the dataset as described in~\cite{strip}, which is technically similar to random poisoning because the dataset is already shuffled before the poisoning. 



\subsection{Threat Model}

We follow a gray-box data poisoning backdoor attack. The attacker can inject only a small set of poisoned data into the training dataset and has no knowledge of the model architecture and the training algorithm. This threat model is realistic as modern datasets are usually based on crowdsourcing both for their creation and validation~\cite{mozillas-common-voice,speech-accent-archive}. 
Thus, an adversary could embed malicious data that could evade any data validation scheme~\cite{large-dataset-pyrrhic-win}.
Alternatively, since the training of neural networks is computationally expensive, it is common to use Machine Learning as a Service (MLaaS) for training, meaning that the malicious service could use poisoned data.

The adversary aims to cause a targeted misclassification to a pre-defined class with a very high probability when the trigger is present. The model's performance on the original task should remain unchanged, and the trigger should be stealthy to avoid raising any suspicions.
Finally, for settings where the signal's sampling rate is lower than 42kHz, the attack would be possible only if we consider a different threat model. Then, the adversary needs to modify the whole training dataset, resulting in a white-box attack regarding data access and model architecture.


\section{Experimental Setup}
\label{sec:setup}


\subsection{Dataset and Input Features}
\label{ssec:dataset}

We use two different versions of the Speech Commands dataset. 
In the first version, we used only the ten classes that were also used in~\cite{meta-neural}. We discarded the files that lasted less than one second to avoid variable-length inputs in our pipeline, resulting in 21\,312 files in total. In the second version, we used the full dataset (30 classes) and discarded the short audio files resulting in 58\,252 files. 

As we discussed in~\Cref{ssec:trigger}, the trigger's sampling rate is 44.1 kHz. 
However, the dataset's audio files are sampled at 16 kHz, so it is impossible to mix the trigger with elements from the dataset. For that reason, we up-sampled the dataset's audio files from 16 kHz to 44.1 kHz. 
This is not a typical sampling rate for speech recognition systems~\cite{google-speech-to-text}, but still supported by popular public APIs~\cite{google-speech-to-text,microsoft-speech-to-text}.


We use the MFCCs as input features.
In particular, we used 40 mel-bands, a step of 10ms (441 samples), and a window length of 25ms (1\,103 samples) which is very common in related works~\cite{generalized-loss-for-speaker-verification,backdoor-attack-against-speaker-verification}. 

\subsection{Neural Network Architectures}

We use two CNNs and one LSTM. The first CNN was also used in~\cite{adversarial-detection-by-classification}. However, we added two dropout layers, one in the penultimate layer (40\%) and one right before the flatten layer (50\%), that gave a performance boost of around 3\%. 
%
%
The second CNN is deeper and was used in~\cite{trojaning-attack-on-nns}, and the LSTM was introduced in~\cite{lstm-attention}. 

All the models perform similarly as the loss function is around 0.3 when the training ends. However, their capacity is different as the first CNN has 284\,778 trainable parameters, the second CNN has 5\,975\,882 trainable parameters, and the LSTM has 180\,569 trainable parameters. In this way, we can investigate whether both simple and more complex networks are susceptible to inaudible backdoor attacks. 
We also investigate whether the backdoor attack behaves differently for different neural networks (CNNs and LSTMs).
We use the Adam optimizer with a learning rate of 0.0001 to train our models and $L_2$ regularization in every convolution filter in both CNNs.
80\% of the initial dataset was used for training/validation, and the remaining 20\% for testing. The training dataset is again split 80\%/20\% for training and validation.

After carefully examining the training and validation loss functions, we saw that the CNNs could always converge before 300 epochs and the LSTM before 100. 
We used Tensorflow's early stopping callback with patience 20, which stops the training if the validation loss is not improved for 20 epochs and returns the weights of the best-performing model.

The arithmetic mean of the epochs among all the experiments we run is 220.8 ($\pm$ 24.97) for the small CNN, 182.4 ($\pm$ 31.39) for the large CNN, and 70.9 ($\pm$ 12.70) for the LSTM for the ten classes of the dataset and 250.5 ($\pm$ 32.96), 102.5 ($\pm$ 29.46), and 83.3 ($\pm$ 13.16) for the 30 classes.
We run each experiment 15 times to eliminate any randomness introduced by the algorithmic randomness (e.g., random initialization and stochastic gradient descent). All the models were trained with Tensorflow 2.5 on NVIDIA RTX 2080 Ti.
\section{Experimental Results}
\label{sec:results}

This section presents a systematic evaluation of our inaudible backdoor attack.
We investigate how the ASRT is affected by various backdoor characteristics like the number of poisoned samples, the trigger's duration, and the trigger's position in time. 

The triggers that we tried can be split into two groups: the long (250ms, 500ms, 750ms, and 1\,000ms) and the short triggers (20ms, 40ms, 60ms, and 80ms). 
In this way, we aim to see if the attack is successful, even if only a few MFCC frames are affected by our trigger. Additionally, we applied our trigger in different positions to see if our models give more attention to the windows that usually contain speech.

\begin{figure}[!ht]
    \centering
    \includegraphics[width=0.48\textwidth]{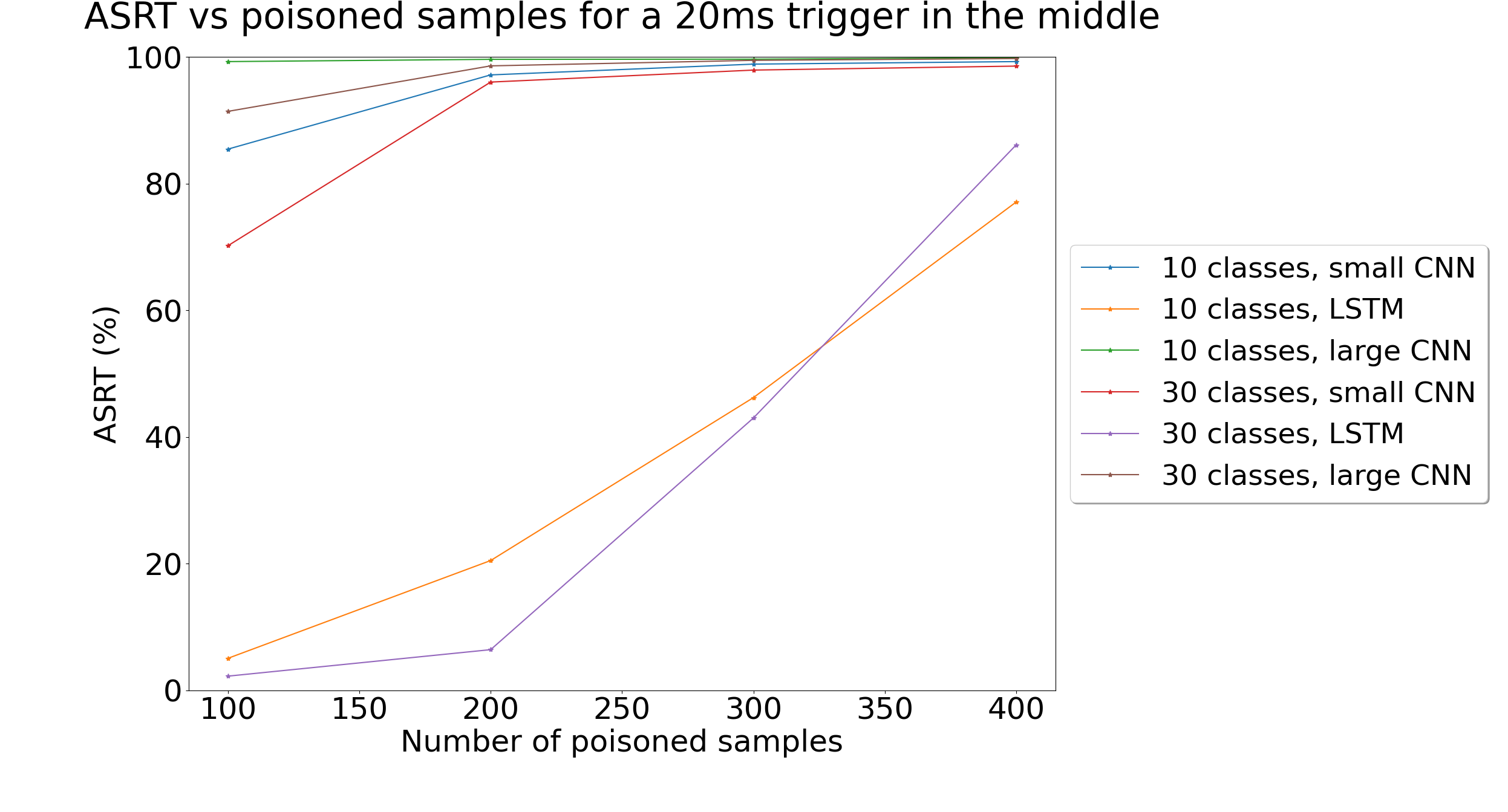}
    \caption{ASRT vs. trigger samples for more than 80 poisoned samples and CNNs.}
    \label{fig:many-samples}
\end{figure}

\begin{figure}
    \centering
    \includegraphics[width=0.40\textwidth]{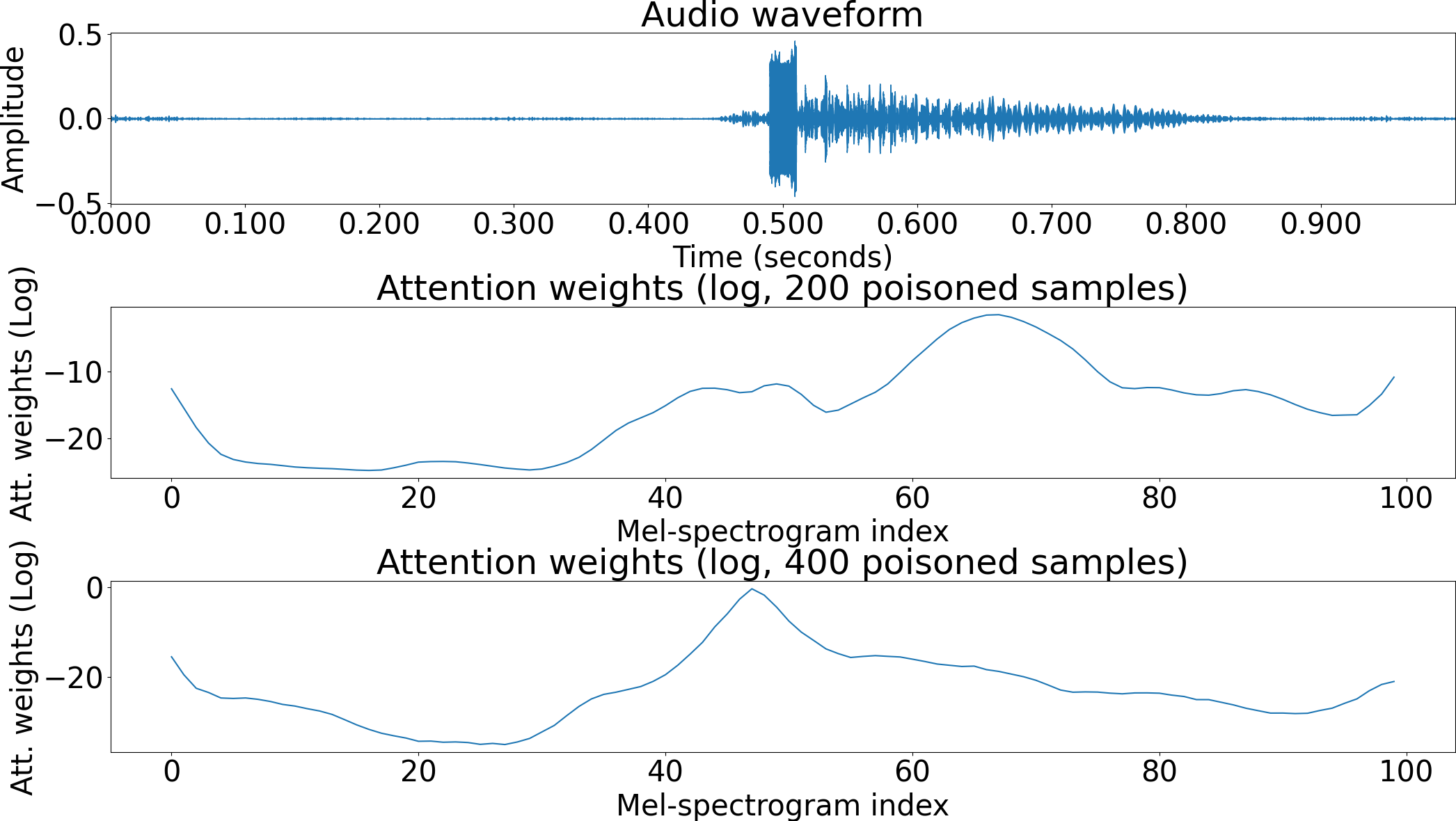}
    \caption{Attention weights for poisoned input ("on").}
    \label{fig:att-weights}
\end{figure}

First, we need to define the poison percentage that leads to an effective attack. \autoref{fig:many-samples} shows ASRT as a function of the poison percentage for each architecture. In these experiments, we used a trigger of 20ms in the middle of each signal which is short and makes the attack very challenging. When we poison 200 samples or more for both CNNs, ASRT is close to optimal even with this short trigger. 
These results are further improved by longer or non-continuous triggers and different trigger positions. Thus, we used a smaller poison percentage to keep any differences between different setups distinguishable. In particular, we poisoned at most 80 (0.59\% for the ten classes and 0.21\% for the 30 classes) training samples in the experiments with the CNNs. The ASRT is lower when we use the LSTM network. 
The attention layer can give useful insights about a network's choices~\cite{lstm-attention}. In~\autoref{fig:att-weights}, we show the logarithm of the attention weights for two backdoored models (20ms trigger in the middle used for the backdoor) when a poisoned input is used. We see that the attention layer does not recognize the trigger when 200 poisoned samples are used. This indicates that the attention layer is more robust to the backdoor attack suggesting that unless a smart position/duration/type of a trigger is used the backdoor attack on LSTM could be rather difficult.

\emph{Roadmap}: In our results, we show ASRT vs. poisoning rate. Each bar chart consists of three subplots, one for every trigger position. Each bar in these plots represents a trigger of a different duration.
Furthermore, to better view the attack effectiveness for different trigger positions, we plot the average ASRT for a different number of poisoned samples in each trigger position. Each of these graphs consists of four lines, one for each position (three lines) and one for the non-continuous trigger. 
We plot only the results for the 30 classes as there are no significant differences between the two versions of the dataset for each architecture.

\subsection{Effect on Clean Accuracy}

A backdoor attack should remain hidden to avoid raising any suspicions. Thus, we need to verify that our attack does not affect the model's performance on the original task. In~\Cref{tab:original-cnn,tab:original-lstm}, we compare the clean model's accuracy on the original task with the accuracy that backdoored models have. To obtain the values for the poisoned models, we group them based on the number of poisoned samples used and take the average (arithmetic mean) of their clean accuracy for all the different triggers used. 

The differences for the original task are negligible between the clean and the backdoored CNNs (\autoref{tab:original-cnn}). The backdoored models perform slightly better in the original task when we use ten classes. 
This behavior indicates that backdoored training data could improve the model's generalization, i.e., backdoored data serves as a regularization factor. However, the differences are small, and further investigation is required to confirm our observations. 
When we use the full dataset, the backdoor insertion drops the clean accuracy in all cases for both CNNs (somewhat larger accuracy drop for large CNN). Again, this is aligned with our previous observations, as more classes (and larger dataset) also represent a more difficult task for neural network architectures to learn.


\begin{table}[!ht]
    \centering
    \small
    \caption{Clean accuracy comparison for the original task for clean and poisoned models (CNNs).}
    \resizebox{0.46\textwidth}{!}{%
        \begin{tabular}{|c|c|c||c|c|c|c|}
        \hline
        &  &  & \multicolumn{4}{c|}{Number of poisoned samples} \\ \hline
        model & classes & original & 20 & 40 & 60 & 80      \\ \hline
        small CNN & 10 & 90.13 ($\pm$ 0.533) & 90.22 ($\pm$ 0.127) & 90.19 ($\pm$ 0.086) & 90.16 ($\pm$ 0.108) & 90.14 ($\pm$ 0.091) \\ \hline
        small CNN & 30 & 87.58 ($\pm$ 0.446) & 87.52 ($\pm$ 0.114) & 87.46 ($\pm$  0.161) & 87.49 ($\pm$ 0.083) & 87.47 ($\pm$ 0.118) \\ \hline
        large CNN & 10 & 95.71 ($\pm$ 0.424) & 95.82 ($\pm$ 0.122) & 95.82 ($\pm$ 0.139) & 95.85 ($\pm$ 0.133) & 95.85 ($\pm$ 0.112) \\ \hline
        large CNN & 30 & 95.14 ($\pm$ 0.539) & 94.82 ($\pm$ 1.527) & 94.61 ($\pm$ 2.609) & 94.24 ($\pm$ 2.292) & 93.52 ($\pm$ 3.621) \\ \hline
        \end{tabular}
    }
    \label{tab:original-cnn}
\end{table}

The LSTM behaves oppositely. In~\autoref{tab:original-lstm}, the clean accuracy is dropped in all cases when the ten classes are used and increased when the entire dataset is used. This indicates that the LSTM built different models in each case and it utilized the model capacity better for ten classes setting. 


\begin{table}[!ht]
    \centering
    \small
    \caption{Clean accuracy comparison for the original task for clean and poisoned models (LSTM).}
    \resizebox{0.46\textwidth}{!}{%
        \begin{tabular}{|c|c|c||c|c|c|c|}
        \hline
        &  &  & \multicolumn{4}{c|}{Number of poisoned samples } \\ \hline
        model & classes & original & 40 & 80 & 120 & 160  \\ \hline
        LSTM & 10 & 91.27 ($\pm$ 1.217) & 91.09 ($\pm$ 0.354) & 91.13 ($\pm$ 0.29) & 90.97 ($\pm$ 0.459) & 90.72 ($\pm$ 0.359) \\ \hline
        LSTM & 30 & 91.45 ($\pm$ 0.748) & 91.76 ($\pm$ 0.198) & 91.63 ($\pm$ 0.208) & 91.59 ($\pm$ 0.24) & 91.77 ($\pm$ 0.241) \\ \hline
        \end{tabular}
    }
    \label{tab:original-lstm}
\end{table}

\subsection{Effect of Trigger Duration, Position, and Continuity on ASRT}

\subsubsection{CNNs}

\begin{figure}[!ht]
    \centering
    \includegraphics[width=0.45\textwidth]{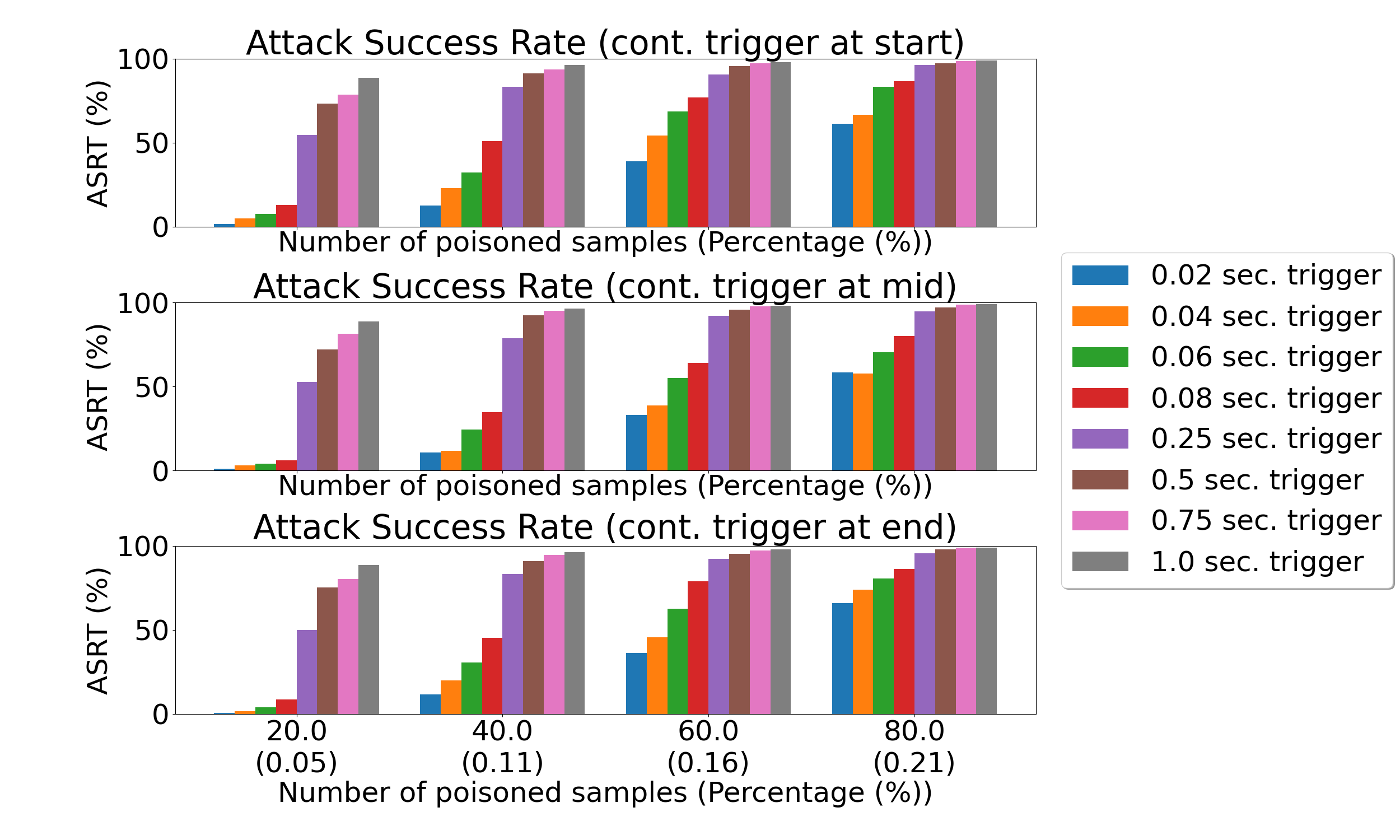}
    \caption{ASRT vs. poisoned samples for all the trigger positions and sizes for 30 dataset classes (small CNN).}
    \label{fig:res-adv-30}
\end{figure}

In~\Cref{fig:res-adv-30}, we show the results for different triggers for the small CNN and the full dataset. The attack is more successful for the large CNN as it is a deeper architecture and is more capable of learning features from only a few training samples. However, we observe some similarities in the attack's behavior in both architectures, and for that reason, we omit the corresponding graph for the large CNN. In particular, ASRT increases linearly as the trigger's duration increases, especially in cases when it is not already high. 
We also see that long triggers result in a high ASRT even if only 20 poisoned samples are used. 
Additionally, ASRT for short triggers is low when using only 20 poisoned samples but is significantly larger when using 80 samples. This shows a strong connection between ASRT and the number of poisoned samples used for the attack.

\begin{figure}[!ht]
    \centering
    \includegraphics[width=0.41\textwidth]{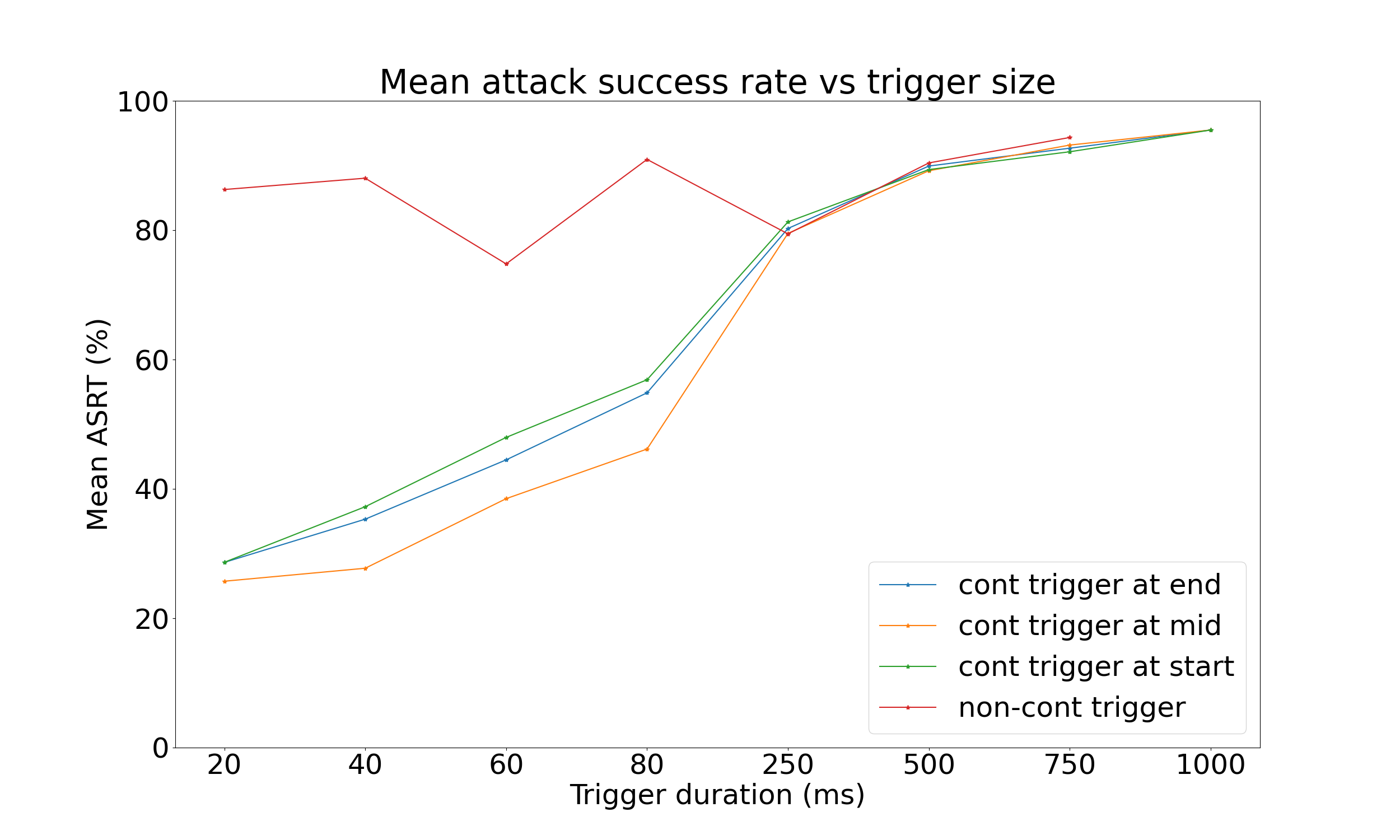}
    \caption{Mean ASRT vs. trigger size for 30 dataset classes (small CNN). 
    }
    \label{fig:res-adv-mean-30}
\end{figure}

In~\Cref{fig:res-adv-mean-30}, we show the mean ASRT for different backdoored models for the small CNN when the full dataset is used. Again, we omit the corresponding graph for the large CNN.
We saw in both CNNs that the position is not important when a long trigger is used. However, the attack effectiveness varies for small triggers. For the small CNN, the non-continuous trigger is substantially more effective than the continuous ones, and the middle of the signal is the worst position for the trigger. The attack is more effective when more MFCC windows are affected, and the trigger does not overlap with actual speech. In the large CNN, the non-continuous trigger outperforms the continuous ones only when the trigger lasts for 20ms. The network's complexity allows effective backdoor insertion even with short triggers longer than 20ms.

\subsubsection{LSTM}

\begin{figure}[!ht]
    \centering
    \includegraphics[width=0.45\textwidth]{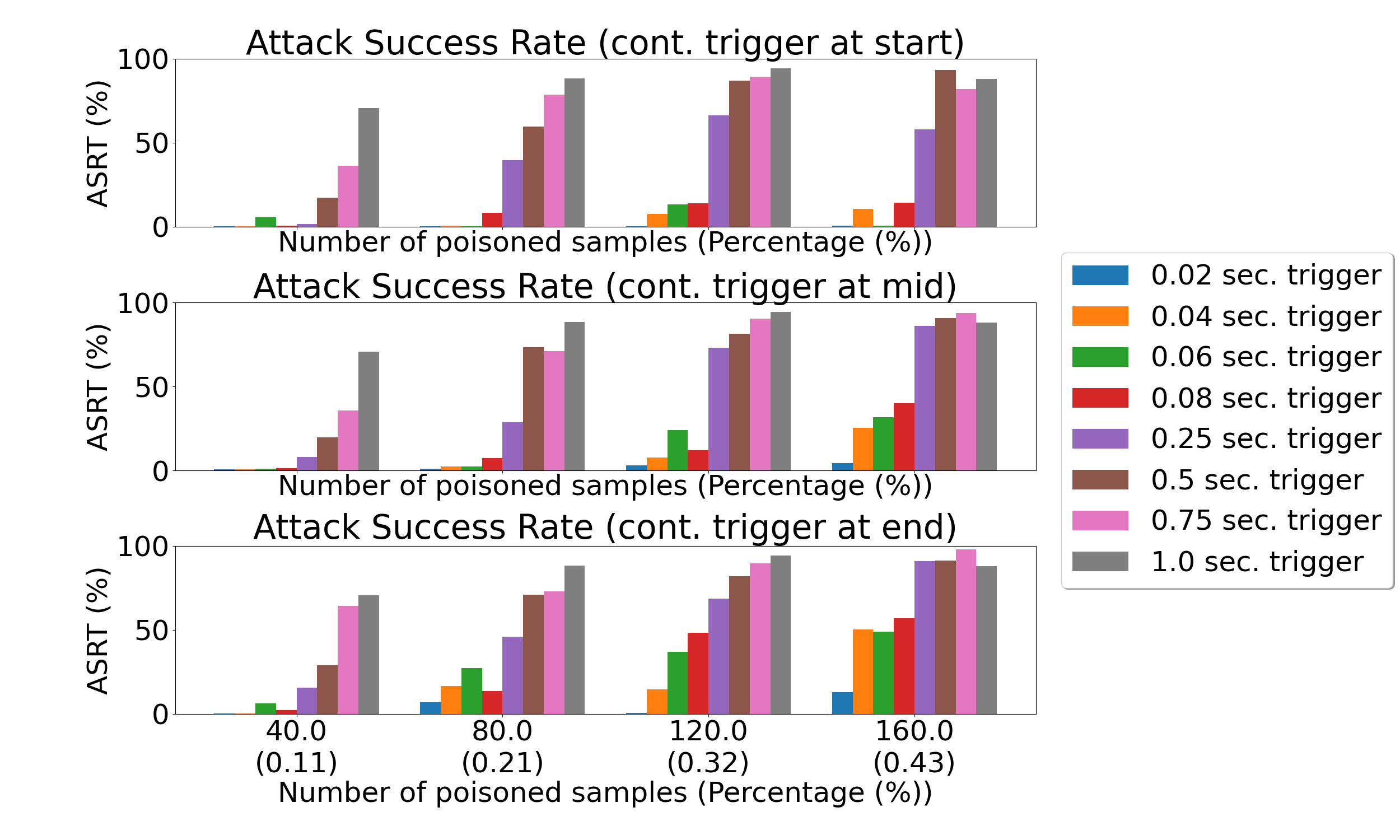}
    \caption{ASRT vs. poisoned samples for all the trigger positions and sizes for 30 dataset classes (LSTM).}
    \label{fig:res-lstm-30}
\end{figure}

In~\autoref{fig:res-lstm-30}, we show the results for each different trigger for the LSTM network for 30 classes. The ASRT is lower for this architecture even though we use more poisoned samples. As shown above, the attention layer cannot be affected easily by small variations inside the audio file.
However, similarly to CNNs, there is a linear correlation between the trigger's duration and ASRT. Again, for long triggers, we see less difference in ASRT, especially when using more poisoned samples.

\begin{figure}[!ht]
    \centering
    \includegraphics[width=0.41\textwidth]{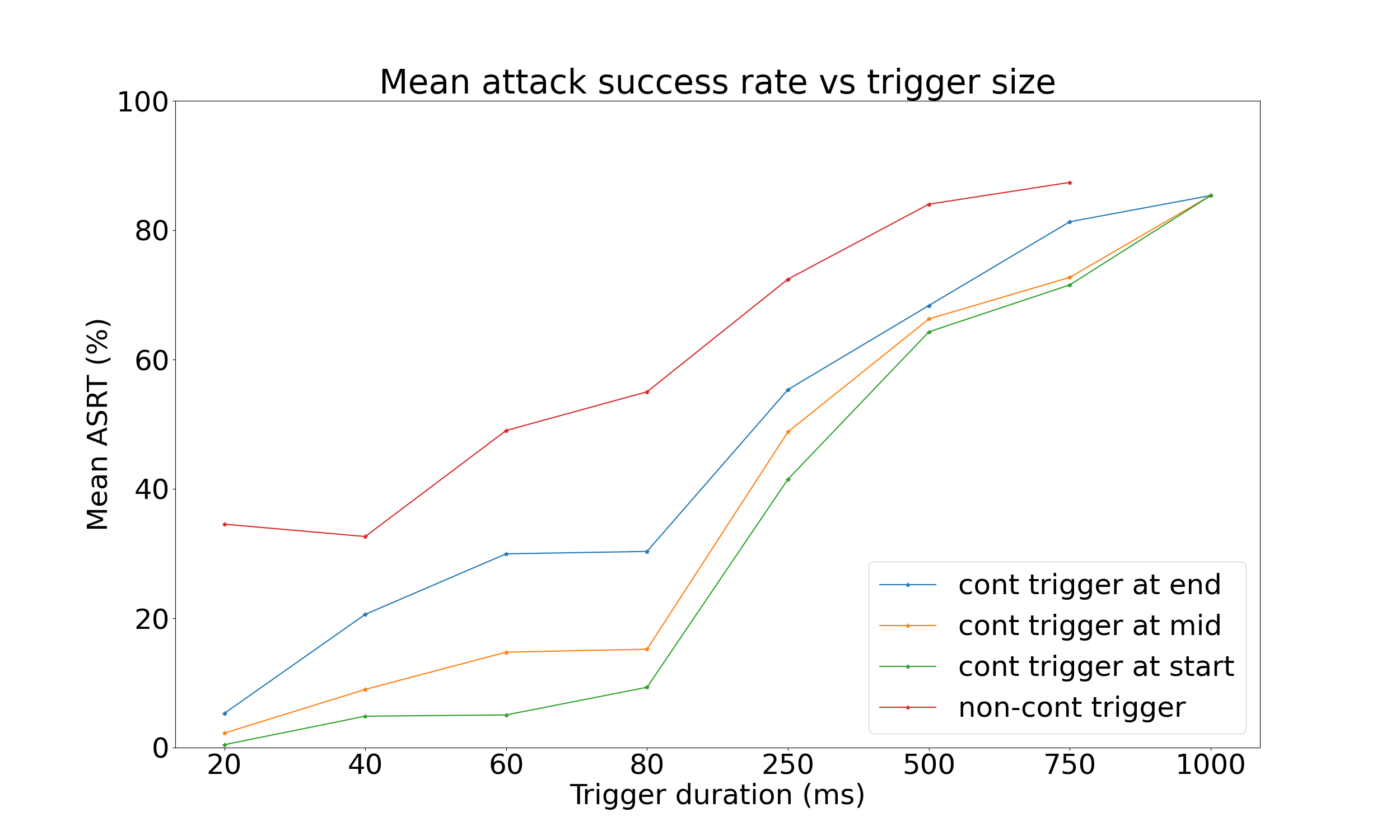}
    \caption{Mean ASRT vs. trigger size for 30 dataset classes (LSTM). 
    }
    \label{fig:res-lstm-mean-30}
\end{figure}

In~\autoref{fig:res-lstm-mean-30}, we show the mean ASRT among different backdoored models for the LSTM network when the full dataset is used. We see that the trigger at the end of the signal is the most effective continuous trigger in both dataset versions. Additionally, non-continuous triggers clearly outperform continuous ones.



\subsection{Attack on Android Application}

Many standard microphones are inherently capable of recording audio signals in the near ultrasonic range (e.g., 18 to 22 kHz)~\cite{backdoor-making-microphones-hear-inaudible-sounds}. 
To this end, we played our trigger through a VLC media player on a Linux laptop and verified through Spectroid~\cite{spectroid} that a signal of 21kHz was played even if we could not hear anything.
This experiment showed that the laptop speaker could transmit sounds in the inaudible human range, and the smartphone microphone could interpret them. Additionally, we verified that the inaudible trigger was successfully generated, which is the first step toward our attack being practical.

As a second step, we applied our attack on an Android speech recognition application. Our application is based on a TensorFlow Lite official example~\cite{android-tf} that recognizes ten spoken words from the Speech Commands dataset. 
However, we modified its preprocessing pipeline to calculate the audio's MFCCs and used a poisoned model for the inference.
We only run this experiment with the small CNN, but we believe this attack will be effective with all three architectures. We decided to use the small CNN because it uses less memory, while the ASRT with the LSTM showed worse performance, making it less attractive architecture.
We used our strongest setup, i.e., 80 poisoned samples and a 1-second trigger to model our adversary. The application records 1-second clips and shows the keywords that are recognized with 50\% probability or more.

We installed our application on a smartphone that ran Android 7.0 (Huawei P9 Lite).
We played our inaudible trigger both from a Linux laptop (Lenovo t460s) and low-end home cinema speakers (Philips HTS7200). 
We run two different experiments for each sound device. In the first experiment, we played our trigger in silence to ensure that the backdoor was successfully activated when no sounds were present. In the second experiment, we said words of the nine possible classes (we did not say the target class) in random order while the trigger was played by either the laptop or the speakers. In each experiment, we placed the smartphone in four different locations to see how the attack accuracy is related to the distance from the sound source. The first location is adjacent to the sound source (0m), and the remaining ones are 0.5m, 1m, and 1.5m away from it. We run each experiment 15 times to have a more robust view of the attack's behavior. 

Using Spectroid~\cite{spectroid}, we saw that for both devices, the trigger's intensity is around $-43$dB when the phone is adjacent to the sound source. However, as the distance between the speaker and the phone grows, this number drops significantly faster for the laptop experiments. Additionally, the trigger's intensity is lower than the normal voice for all the positions we tried. As a result, the backdoor was successfully triggered only in silence as its volume was not enough to overcome our speech. In particular, ASRT is high only when the distance is 0m if the trigger is played through the laptop. On the other hand, if the trigger is played from the home cinema speakers, ASRT is high ($>80\%$) for distances up to 1m. We saw that ASRT is further increased when the user does not speak loud enough or the phone is not close enough to the speaker's mouth. We concluded from these experiments that this attack poses a real threat as an adversary with stronger equipment could be more effective.

\subsection{General Observations}

Based on the conducted experiments, we list several findings that were consistently observed.
\begin{compactitem}
\item The attack success rate is linearly correlated with the trigger's duration. Our models can learn longer triggers easier as more windows in the MFCC calculation are affected. 
\item Our attack is inaudible, so an adversary could use a 1-second trigger for a powerful attack with a very low poisoning rate.
\item If the adversary does not want to use a 1-second trigger (or any long trigger), non-continuous triggers represent the best option as they perform extremely well with short triggers.
\item  We saw that different models behave differently for our attack. For example, the best ASRT for the short continuous triggers and LSTM is at the end of the signal, but for the large CNN, the trigger at the beginning results in a slightly increased ASRT. This indicates that the data poisoning backdoor attack could be used as a tool for AI explainability, as it can highlight how and what neural networks learn.
\item The number of poisoned samples is clearly connected with ASRT. Thus, an adversary should use as many samples as possible without risking a large performance drop for the original task to avoid raising suspicions.
\item The backdoor in the Android application was successfully activated by playing the trigger from the laptop and low-end home cinema speakers within a 1.5m.
\end{compactitem}




\subsubsection*{Limitations}

While our attack showed very good performance, we believe there are also some limitations concerning inaudible backdoor attacks:
\begin{compactitem}
\item We use MFCCS as a convenient representation of audio signals for machine learning, and we did not validate our results with other types of signal representations. Still, we do not consider this a significant limitation due to the widespread use of MFCCS and no intuitive reason that the attack would not work with different representations.
\item As we work with inaudible triggers, the sampling rate of the recorded signals should be larger than 40 kHz.
If an ASR system uses a lower sampling rate, our attack will not be effective. 
A sampling rate of 16kHz is preferred in ASR systems~\cite{google-speech-to-text}, however popular public speech-to-text APIs allow higher rates~\cite{google-speech-to-text,microsoft-speech-to-text}. Additionally, different applications like music genre classification require high sampling rates~\cite{dnn-music-genre-classification}, making our attack practical.
\item Finally, our work did not consider any defense against the inaudible backdoor attack. While we show it is possible to achieve (almost) 100\% ASRT for various trigger positions and durations even if using a small number of poisoned examples, different defenses could still work. Our trigger works for every source class in the dataset, and thus, an online defense like STRIP~\cite{strip} could defend from it by superimposing different signals and checking the entropy of the model's output. Additionally, a low-pass filter could make our attack ineffective as it could filter out frequencies above the human audible range. Such a filter could be either software-based or hardware-based. A software-based filter could introduce additional computational overhead, which may be unacceptable in a device with strict timing constraints. Moreover, a hardware-based filter may not be possible in any device as frequencies close to the ultrasonic range may be useful in some cases like music genre classification~\cite{dnn-music-genre-classification}.
\item When playing a poisoned audio sample, if the volume is too high, it is possible to hear a slight clicking sound in its beginning. This clicking sound is connected with the backdoor trigger. Its volume is very low compared to the main signal, so unless expecting it, it is hard to hear it. This sound is heard only in the trigger's beginning and thus, we still consider it to be practically inaudible.
\end{compactitem}


\section{Conclusions and Future Work}
\label{sec:conclusions}

This paper explores how inaudible backdoor attacks on neural networks threaten automatic speech recognition systems. 
We use a dataset with ten or 30 classes and three neural networks (two CNNs and one LSTM). Furthermore, we investigate the influence of the position, duration, and trigger type.
Our results suggest it is relatively easy to run an inaudible backdoor attack on ASR where the attacker needs to poison only around 0.5\% of the training dataset. Since the trigger is inaudible, it is possible to make it as long as the signal, making the attack even more powerful. We show that non-continuous triggers can significantly improve attack performance even in scenarios where short triggers are used. Consequently, short non-continuous triggers inserted in less than 0.5\% of the training dataset can result in an attack success rate of more than 99\%. Finally, we show that our attack is effective in the real-word by attacking an Android application.

As we did not consider countermeasures, this would be a natural extension of our work. Today, most of the defenses against backdoor attacks try to detect either the poisoned inputs or compromised models~\cite{DBLP:journals/corr/abs-1804-00308,DBLP:journals/corr/abs-1811-03728}. As we limit our data poisoning to a small number of training examples, we believe such techniques would have difficulties detecting poisoning examples. Furthermore, some more recent defenses suggested retraining~\cite{DBLP:journals/corr/abs-2004-11514}, which we consider to be a good choice against our attack. Still, since such works commonly assume an unlimited amount of clean data, they are not practical. We are aware of only one approach that does not have that limitation~\cite{DBLP:journals/corr/abs-2007-00711}, but it can work for image processing only, making this defense not applicable against our attack.

\balance

\bibliographystyle{ACM-Reference-Format}
\bibliography{main}


\end{document}